\documentstyle[12pt]{article}

\begin{document}
\begin{titlepage}
\title{ \Large \bf Colliding Gravitational Plane Waves in Dilaton Gravity }
\vspace{5cm}
\author{Metin G{\" u}rses and Emre Sermutlu\\
{\small Department of Mathematics, Faculty of Science} \\
{\small Bilkent University, 06533 Ankara - Turkey}\\
e-mail: gurses@fen.bilkent.edu.tr}
\maketitle
\begin{abstract}
Collision of plane waves in dilaton gravity theories and low energy limit
of string theory is considered. The formulation of the problem
and some exact solutions are presented
\end{abstract}

\end{titlepage}
\section{Introduction}

Plane wave geometries are not only important in classical general
relativity but also in string theory .
It is now very well known
that these geometries are the exact classical solutions of the
string theory at all orders of string tension parameter
\cite{GUV}-\cite{HOW}. It is
also interesting that plane wave metrics in higher dimensions
when dimensionaly reduced lead to exact extreme black hole
solution in string theory \cite{HOR}.

In this work we shall be interested
in the head on collisions of these plane waves in the framework
of Einstein-Maxwell-Dilaton theories with one $U(1)$ and two
$U(1)$ abelian gauge fields \cite{KAL}. Our formulation of the problem
will also cover the low energy limit of the string theory
for some fixed values of the dilaton coupling constants.
Hence the solutions we present in this work are also exact solutions
of the low energy limit of string theories. We give the complete data for
the colliding plane-shock waves.
We formulate the collision of plane waves and give solutions
for the colinear case. When the dilaton coupling constant vanishes
one of our solutions reduces to the well known Bell-Szekeres
solution \cite{BEL} in Einstein Maxwell Theory.

For the collision problem in general relativity spacetime is divided
into four regions with respect to the null coordinates u and v.
The second and third (incoming) regions are the Cauchy data
(characteristic initial data) for the field equations in the
interaction region (IV. Region). For this purpose the specification of the
data is quite important in the formulation of the collision problem
\cite{kp}-\cite{GRI}.
We show that the future closing singularities appearing
in classical solutions exist also in dilaton gravity and in the low energy
limit of the string theory. This is due to focusing effect of the plane waves
\cite{PEN}.

It is an
open question whether this classical treatment of collision of plane waves
can be extended to all orders in the string tension parameter \cite{PAP},
\cite{ARE}. One of the limiting cases of the solutions in section 2 is the
Bell-Szekeres solution \cite{BEL}. This solution seems to be a canditate
for an exact solution at all orders.
The Bell-Szekeres solution in the
interaction region is diffeomorphic to the Bertotti-Robinson spacetime
\cite{HAY} ,\cite{GRI}. It is known that string theory preserves the
form of Bertotti-Robinson metric at all orders of the string parameter
\cite{GU1} , \cite{STR}. This does not necessarily lead to a conclusion
that the Bell-Szkeres solution is an exact solution of the string theory.
The reason is that the diffeomorphism is valid only in the interaction region
($u>0 , v>0$) and hence the field equations
(at higher orders of the string parameter)
may not be satisfied on the hyperplanes $u=0$ and $v=0$.
The Weyl tensor and its covariant derivatives suffer from  delta
function and derivatives of the delta function type of
singularities on the hyperplanes $u=0$ and $v=0$.
It is unlikely that these singular terms cancel each
other in the field equations
at all orders. If there exists an exact solution reperesenting
the collision of plane waves in the full string theory
then its low energy limit should be contained in
our solutions in the second and third sections. The proof of this conjecture
is of course not easy to prove.

In the next sections we shall give the form of the metrics
in the incoming regions. These will constitute the data for field equations
in the interaction region.
In the second section we give the formulation of the problem for one
$U(1)$ abelian gauge field with a solution generalizing the Bell-Szekeres
solution in general relativity. In the third section we consider
two abelian $U(1)$ gauge fields and give some interesting exact solutions
of the collision of plane waves problem. In the Appendix we reduce the
Maxwell Dilaton field equations , in the collision of plane waves , to two
dimensional Ernst equation.

\section{\bf Dilaton gravity with one $U(1)$ vector field}

\noindent
Einstein-Maxwell-Dilaton Gravity is derivable from a variational principle
with the lagrangian density

\begin{equation}
L={\sqrt{ -g}}\left[{R\over 2\kappa^{2}}-
{2\over \kappa^{2}}(\nabla \psi)^{2}-{1\over 4}\ e^{-a \psi}F^{2}\right].
\end{equation}

\noindent
where $a$ is the dilaton coupling constant. The field equations are

\begin{equation}
G_{\mu \nu}=4\,\left[\partial_{\mu} \psi\,\partial_{\nu}\psi-
{1\over 2}(\nabla\psi)^{2} g_{\mu \nu} \right] +
\kappa^{2} e^{-a \psi}\left[F_{\mu}^{\alpha} F_{\nu \alpha}-{1\over 4} F^{2}
g_{\mu \nu} \right], \label{EJ}
\end{equation}

\begin{equation}
\nabla_{\mu} (e^{-a \psi } F^{\mu \nu}) = 0,  \label{EI}
\end{equation}

\begin{equation}
\partial_{\mu} ({\sqrt -g}\, g^{\mu \nu}\, \partial_{\nu} \psi)+
{\kappa^{2}\, a\, {\sqrt -g} \over 16} e^{-a \psi} F^2 = 0,  \label{EE}
\end{equation}

\noindent
A space time describing the collision of plane waves admits two
space like Killing vector fields. In the general case these vectors
are nonorthogonal but here in this work we consider them to be orthogonal.
For such a case an appropriate form of
the metric $g_{\mu \nu}$ and $U(1)$ gauge potential $A_{\mu}$ are given by

\begin{equation}
ds^{2}=2e^{-M}du\,dv+e^{-U-V}dy^{2}+e^{-U+V}dz^{2}
\label{me1}
\end{equation}

\begin{equation}
A_{\mu}=(0,0,A,0)
\end{equation}

\noindent
where $M=M(u,v)$, $U=U(u,v)$, $V=V(u,v)$, $A=A(u,v)$  and dilaton
field $\psi=\psi(u,v)$. The field equations turn out to be

\begin{equation}
-2A_{,uv}=(V_{u}-a\, \psi_{u})\, A_{,v}+(V_{v}-a\, \psi_{v})\, A_{,u}
\label{aik}
\end{equation}

\begin{equation}
U_{uv}-U_{u}U_{v}=0
\end{equation}

\begin{equation}
2M_{uv}=-2\,U_{uv}+U_{u}U_{v}+V_{u}V_{v}+8\psi_{u}       \psi_{v}
\label{obt}
\end{equation}

\begin{equation}
2V_{uv}-U_{u}V_{v}-U_{v}V_{u}-2\kappa^{2} e^{U+V-a\psi} A_{,u} A_{,v} =0
\label{bir}
\end{equation}

\begin{equation}
2\psi_{uv}-U_{u} \psi_{v}   -U_{v} \psi_{u} +{a \kappa^{2} \over 4}
e^{U+V-a\psi} A_{,u} A_{,v}=0 \label{iki}
\end{equation}

\begin{equation}
-2M_{u}U_{u}-2\,U_{uu}+U_{u}^{2}+V_{u}^{2}+8\psi_{u}     ^{2}
+2\kappa^{2} e^{U+V-a\psi} A_{,u}^{2}=0 \label{ana}
\end{equation}

\begin{equation}
-2M_{v}U_{v}-2\,U_{vv}+U_{v}^{2}+V_{v}^{2}+
8\psi_{v}        ^{2}+2\kappa^{2} e^{U+V-a\psi} A_{,v}^{2}=0 \label{anb}
\end{equation}

\noindent
Note that  (\ref{obt}) can be derived from the other equations. It
is not independent.
 From (\ref{bir}) and (\ref{iki}), letting $E=V-a\psi$ we obtain

\begin{equation}
2E_{uv}-U_{u}E_{v}-U_{v}E_{u}-(2+{a^{2}\over 4})\kappa^{2} e^{U+E} A_{,u}
A_{,v} \label{son}
\end{equation}

\noindent
Letting

\begin{equation}
B=\sqrt{2+{a^{2} \over 4}}\, \kappa\,A
\end{equation}

\noindent
(\ref{aik}) and (\ref{son}) become

\begin{equation}
-2\,B_{,uv}=E_{u} B_{,v}+E_{v} B_{,u} \label{s11}
\end{equation}

\begin{equation}
2E_{uv}-U_{u}E_{v}-U_{v}E_{u}- e^{U+E}\,B_{,u}\,B_{,v}=0 \label{s12}
\end{equation}

\noindent
The above two equations are the real and imaginary
parts of the following Ernst equation

\begin{equation}
Re(\varepsilon)\, \nabla^{2}\, \varepsilon =
\nabla \varepsilon\, \nabla \varepsilon  \label{s01}
\end{equation}

\noindent
where differential operators in (\ref{s01}) are defined with
respect to the metric given by $ds^{2}=2\,du\,dv-e^{-2\,U}\,d\, \phi^{2}$
and

\begin{equation}
\varepsilon=e^{-{1 \over 2}\,(E+U)}+i{B \over \sqrt{2}}
\end{equation}

\noindent
The remaining part of the Einstein equations are given as follows

\begin{equation}
U_{uv}-U_{u}U_{v}=0 \label{s13}
\end{equation}

\begin{equation}
2\,X_{uv}-U_{u}X_{v}-U_{v}X_{u} =0 \label{s14}
\end{equation}

\begin{equation}
-2M_{u}U_{u}-2\,U_{uu}+U_{u}^{2}+{1 \over \alpha}\,(E_{u}^{2}+8\,X_{u}^{2})
+2\kappa^{2} e^{U+E} A_{,u}^{2}=0 \label{s15}
\end{equation}

\begin{equation}
-2M_{v}U_{v}-2\,U_{vv}+U_{v}^{2}+{1 \over \alpha}\,(E_{v}^{2}+
8\,X_{v} ^{2})+2\kappa^{2} e^{U+E} A_{,v}^{2}=0 \label{s16}
\end{equation}

\noindent
where

\begin{equation}
\psi={1 \over \alpha}\,(X-{a \over 8}\,E) ~,~
V={1 \over \alpha}\,(a\,X +\,E)~ , ~
\alpha=1+{a^{2} \over 8} \label{PSI}
\end{equation}

\noindent
Hence a solution of the Dilaton Gravity field equations depends
upon a linear equation (\ref{s14}) and the Ernst equation (\ref{s01}).
The integrability of the Ernst equation and its properties are now
very well known \cite{HON} , but the characteristic
initial value problem has not been solved yet.

\noindent
The formulation of the collision of plane waves is as follows: The space time
is divided
into four disjoint regions by the null hyperplanes $u=0$ and $v=0$

\noindent
{\bf The first region}: ($u \le 0 , v \le 0$)

\begin{equation}
ds^{2}=2du\,dv+dy^{2}+dz^{2} \label{FLT}
\end{equation}

\noindent
This is the flat space time with $\psi=A=0$.

\noindent
{\bf The second region}: ($u > 0 , v \le 0$)

\begin{equation}
ds^{2}=2e^{-M_{2}}du\,dv+e^{-U_{2}-V_{2}}dy^{2}+e^{-U_{2}+V_{2}}dz^{2}
\label{s17}
\end{equation}

\noindent
where $M_{2}=M_{2}(u)$ , $U_{2}=U_{2}(u)$ , $V_{2}=V_{2}(u)$ ,
$\psi_{2}=\psi_{2}(u)$ and $A_{2}=A_{2}(u)$
constitute the data at $v \le 0$. The only field equation is

\begin{equation}
-2M_{2,u}U_{2,u}-2\,U_{2,uu}+U_{2,u}^{2}+{1 \over \alpha}\,
(E_{2,u}^{2}+8\,X_{2,u}^{2})
+2\kappa^{2} \,e^{U_{2}+E_{2}}\,A_{2,u}^{2}=0 \label{s18}
\end{equation}

\noindent
{\bf The third region}: ( $u \le 0 , v > 0$)

\begin{equation}
ds^{2}=2e^{-M_{3}}du\,dv+e^{-U_{3}-V_{3}}dy^{2}+e^{-U_{3}+V_{3}}dz^{2}
\label{s07}
\end{equation}

\noindent
where $M_{3}=M_{3}(v)$ , $U_{3}=U_{3}(v)$ , $V_{3}=V_{3}(v)$ ,
$\psi_{3}=\psi_{3}(v)$ and $A_{3}=A_{3}(v)$
constitute the data at $u \le 0$. The only field equation is

\begin{equation}
-2M_{3,v}U_{3,v}-2\,U_{3,vv}+U_{3,v}^{2}+{1 \over \alpha}\,(E_{3,v}^{2}+
8\,X_{3,v} ^{2})+2\kappa^{2} \,e^{U_{3}+E_{3}}\,A_{3,v}^{2}=0 \label{s19}
\end{equation}

\noindent
The second and third regions are called the incoming regions
and the corresponding space times are the plane wave geometries.
Hence the functions $M_{2}=M_{2}(u)$ , $U_{2}=U_{2}(u)$ , $V_{2}=V(u)$ ,
$\psi_{2}=\psi_{2}(u)$ ,
$A_{2}=A_{2}(u)$ and  $M_{3}=M_{3}(v)$ , $U_{3}=U_{3}(v)$ , $V_{3}=V_{3}(v)$ ,
$\psi_{3}=\psi_{3}(v)$ ,
$A_{3}=A_{3}(v)$ should be considered as the data on the hyperplanes
$v=0$ and $u=0$ respectively.

\noindent
{\bf The fourth region}: ( $u > 0 , v > 0$ )

\noindent
The metric takes the form (\ref{me1})
with $M=M(u,v)$ , $U=U(u,v)$ , $V=V(u,v)$ , $\psi=\psi(u,v)$ and
$A=A(u,v)$ such that in the incoming regions
($u \le 0 , v \le 0$) the metric (\ref{me1}) reduces to the corresponding
metrics in the related regions. The field equations are given in
Eqs.(\ref{s01}) and (\ref{s13})-(\ref{s13}).

\noindent
The problem is to find the solutions of the above equations in
such a way that the following conditions must be satisfied.

\begin{eqnarray}
M(u,v \le 0)=M_{2}(u) , U(u,v \le 0)=U_{2}(u) , V(u,v \le 0)=V_{2}(u)\\
\psi(u,v \le 0)=\psi_{2}(u) , A(u,v \le 0)=A_{2}(u)
\end{eqnarray}

\begin{eqnarray}
M(u \le 0 ,v)=M_{3}(v) , U(u \le 0 ,v)=U_{3}(v) , V(u \le 0 ,v)=V_{3}(v)\\
\psi(u \le 0 ,v)=\psi_{3}(v) , A(u \le 0 ,v)=A_{3}(v)
\end{eqnarray}

\noindent
An exact solution of the above problem is.

\begin{eqnarray}
U&=&-\ln \cos (P+Q)-\ln \cos (P-Q) \nonumber \\
E&=&\ln \cos (P+Q)-\ln \cos (P-Q)  \\
A&=&\rho \sin (P-Q)\\
X&=&{k_{1}\over 2}\, \ln\,{\cos Q - \sin P \over \cos Q + \sin P}+{k_{2}\over
2}\,
\ln\,{\cos P - \sin Q \over \cos P + \sin Q}
\end{eqnarray}

\noindent
Here $P=a_{2}\,u\,\theta(u)$ , $Q=a_{3}\,v\, \theta(v)$ , where
$\theta$ is the Heaviside step function , $a_{2}$ and $a_{3}$ are
arbitrary constants.

\begin{equation}
\rho^{2}={16\over (8+a^{2}) \kappa^{2}}
\end{equation}

\noindent
There are two distinct solutions.

\noindent
{\bf 1.$ k_{1}=k_{2}=k$ and $k^{2}={a^{2} \over 16}$}

\noindent
{\bf Second region}: $v \le 0 , u > 0$ or $Q=0$

\begin{eqnarray}
e^{a\,\psi_{2}}&=&\left[{1-\sin\,P \over 1+\sin\,P}\right]^{{a\,k \over 2\,
\alpha}}\\
e^{-U_{2}-V_{2}}&=&\cos^{2}\, P \,\left[{1-\sin\,P \over
1+\sin\,P}\right]^{-{a\,k \over 2\, \alpha}}\\
e^{-U_{2}+V_{2}}&=&\cos^{2}\, P \,\left[{1-\sin\,P \over
1+\sin\,P}\right]^{{a\,k \over 2\, \alpha}}\\
e^{-M_{2}}&=&(\cos\,P)^{a^{2} \over 8\, \alpha}\\
A_{2}&=&\rho\, \sin\,P
\end{eqnarray}

\noindent
{\bf Third region}: $u \le 0 , v > 0$ or $P=0$.

\begin{eqnarray}
e^{a\,\psi_{3}}=\left[{1-\sin\,Q \over 1+\sin\,Q}\right]^{{a\,k \over 2\,
\alpha}}\\
e^{-U_{3}-V_{3}}=\cos^{2}\, Q \,\left[{1-\sin\,Q \over
1+\sin\,Q}\right]^{-{a\,k \over 2\, \alpha}}\\
e^{-U_{3}+V_{3}}=\cos^{2}\, Q \,\left[{1-\sin\,Q \over
1+\sin\,Q}\right]^{{a\,k \over 2\, \alpha}}\\
e^{-M_{3}}=(\cos\,Q)^{a^{2} \over 8\, \alpha}\\
A_{3}=-\rho\, \sin\,Q
\end{eqnarray}

\noindent
{\bf Fourth region}: $u > 0 , v > 0$ .

\begin{center}
$$e^{a\,\psi}=\left[{(\cos\,Q-\sin\,P)\,(\cos\,P-\sin\,Q) \over
(\cos\,Q+\sin\,P)\,(\cos\,P+\sin\,Q)}\right]^{{a\,k \over 2\, \alpha}}\,
({\cos\,(P-Q) \over \cos(P+Q)})^{a^{2} \over 8\, \alpha}$$
$$e^{-U-V}=(\cos\,(P+Q))^{1-{1 \over  \alpha}}\,
(\cos\,(P-Q))^{1+{1 \over  \alpha}}
 \,\left[{(\cos\,Q-\sin\,P)\,(\cos\,P-\sin\,Q) \over
(\cos\,Q+\sin\,P)\,(\cos\,P+\sin\,Q)}\right]^{-{a\,k \over 2\, \alpha}}$$
$$e^{-U+V}=(\cos\,(P+Q))^{1+{1 \over  \alpha}}\,
(\cos\,(P-Q))^{1-{1 \over  \alpha}}
 \,\left[{(\cos\,Q-\sin\,P)\,(\cos\,P-\sin\,Q) \over
(\cos\,Q+\sin\,P)\,(\cos\,P+\sin\,Q)}\right]^{{a\,k \over 2\, \alpha}}$$
$$e^{-M}=\left(\cos\,(P+Q)\right)^{3\,a^{2} \over 16\, \alpha}\,
\left(\cos\,(P-Q)\right)^{-\,a^{2} \over 16\, \alpha}$$
$$A=\rho\, \sin\,(P-Q)$$
\end{center}

\noindent
{\bf 2.$ k_{2}=-k_{1}=-k$ and $k^{2}={a^{2} \over 16}$}

\noindent
{\bf Second region}: $v \le 0 , u > 0$ or $Q=0$

\begin{eqnarray}
e^{a\,\psi}=\left[{1-\sin\,P \over 1+\sin\,P}\right]^{{a\,k \over 2\,
\alpha}}\\
e^{-U-V}=\cos^{2}\, P \,\left[{1-\sin\,P \over
1+\sin\,P}\right]^{-{a\,k \over 2\, \alpha}}\\
e^{-U+V}=\cos^{2}\, P \,\left[{1-\sin\,P \over
1+\sin\,P}\right]^{{a\,k \over 2\, \alpha}}\\
e^{-M}=(\cos\,P)^{a^{2} \over 8\, \alpha}\\
A_{2}=\rho\, \sin\,P
\end{eqnarray}

\noindent
{\bf Third region}: $u \le 0 , v > 0$ or $P=0$.

\begin{eqnarray}
e^{a\,\psi}=\left[{1+\sin\,Q \over 1-\sin\,Q}\right]^{{a\,k \over 2\,
\alpha}}\\
e^{-U-V}=\cos^{2}\, Q \,\left[{1+\sin\,Q \over
1-\sin\,Q}\right]^{-{a\,k \over 2\, \alpha}}\\
e^{-U+V}=\cos^{2}\, Q \,\left[{1+\sin\,Q \over
1-\sin\,Q}\right]^{{a\,k \over 2\, \alpha}}\\
e^{-M}=(\cos\,Q)^{a^{2} \over 8\, \alpha}\\
A_{2}=-\rho\, \sin\,Q
\end{eqnarray}

\noindent
{\bf Fourth region}: $u > 0 , v > 0$ .

\begin{center}
$$e^{a\,\psi}=\left[{(\cos\,Q-\sin\,P)\,(\cos\,P+\sin\,Q) \over
(\cos\,Q+\sin\,P)\,(\cos\,P-\sin\,Q)}\right]^{{a\,k \over 2\, \alpha}}\,
\left({\cos\,(P-Q) \over \cos(P+Q)}\right)^{a^{2} \over 8\, \alpha}$$\\
$$e^{-U-V}=(\cos\,(P+Q))^{1-{1 \over  \alpha}}\,
(\cos\,(P-Q))^{1+{1 \over  \alpha}}
 \,\left[{(\cos\,Q-\sin\,P)\,(\cos\,P+\sin\,Q) \over
(\cos\,Q+\sin\,P)\,(\cos\,P-\sin\,Q)}\right]^{-{a\,k \over 2\, \alpha}}$$\\
$$e^{-U+V}=(\cos\,(P+Q))^{1+{1 \over  \alpha}}\,
(\cos\,(P-Q))^{1-{1 \over  \alpha}}
 \,\left[{(\cos\,Q-\sin\,P)\,(\cos\,P+\sin\,Q) \over
(\cos\,Q+\sin\,P)\,(\cos\,P-\sin\,Q)}\right]^{{a\,k \over 2\, \alpha}}$$\\
$$e^{-M}=\left(\cos\,(P-Q)\right)^{3\,a^{2} \over 16\, \alpha}\,
\left(\cos\,(P+Q)\right)^{-\,a^{2} \over 16\, \alpha}$$\\
$$A=\rho\, \sin\,(P-Q)$$
\end{center}

\noindent
The spacetime in the fourth region is singular
on the hyperplanes $a_{2}\,u \pm a_{3}\,v={\pi \over 2}$. When $a$ goes to
zero both of the above solutions reduce to the well known Bell-Szekeres
solution \cite{BEL}.

\section{\bf Dilaton gravity with two $U(1)$ vector fields}

A dimensionally reduced superstring theory in four
dimensions can be described in terms of $N=4$ supergravity \cite{KAL}.
There are two versions $N=4$ supergravity , SO(4) and SU(4) versions.
We shall only consider bosonic part
of the theory with $U(1)\bigotimes U(1)$ vectors in each version
and one real dilaton field. In the following lagrangian although
$(a,b)=(2,-2)$ for $SO(4)$ case and $(a,b)=(2,2)$ for $SU(4)$ case
we shall keep these constants (couplings of dilaton field to each
gauge field)

\begin{equation}
L={\sqrt{ -g}}\left[{R\over 2\kappa^{2}}-
{2\over \kappa^{2}}(\nabla \psi)^{2}-
{1\over 4}\,( e^{-a \psi}\,F^{2}+ e^{-b \psi}\,H^{2})\right].
\end{equation}

\noindent
The field equations are

\begin{eqnarray*}
G_{\mu \nu}=4\,\left[\partial_{\mu} \psi\,\partial_{\nu}\psi-
{1\over 2}(\nabla\psi)^{2} g_{\mu \nu} \right] +
\kappa^{2}\, e^{-a \psi}\left[F_{\mu}^{\alpha} F_{\nu \alpha}-
{1\over 4} F^{2} g_{\mu \nu} \right] +\\ \kappa^{2}\,
e^{-b \psi}\left[H_{\mu}^{\alpha} H_{\nu \alpha}-
{1\over 4} H^{2} g_{\mu \nu} \right], \label{EEJ}
\end{eqnarray*}

\begin{equation}
\nabla_{\mu} (e^{-a \psi } F^{\mu \nu}) = 0,  \label{EEI}
\end{equation}

\begin{equation}
\nabla_{\mu} (e^{-b \psi } H^{\mu \nu}) = 0,  \label{EE2}
\end{equation}

\begin{equation}
\partial_{\mu} ({\sqrt -g}\, g^{\mu \nu}\, \partial_{\nu} \psi)+
{\kappa^{2}\, {\sqrt -g} \over 16} (a\,e^{-a \psi} F^2\,+ b\,
e^{-b \psi} H^2)=0,  \label{EEE}
\end{equation}

\noindent
where $F^{2}=F^{\alpha \, \beta}\, F_{\alpha \, \beta}$ and
$H^{2}=H^{\alpha \, \beta}\, H_{\alpha \, \beta}$ . Both
$F_{\mu \nu}$ and $H_{\mu \nu}$ are obtained by the vector
potentials $A_{\mu}$ and $B_{\mu}$ respectively , i.e.,
they are given by

\begin{equation}
F_{\mu \nu}= \partial_{\mu}\,A_{\nu} - \partial_{\nu}\,A_{\mu} ~~,~~
H_{\mu \nu}= \partial_{\mu}\,B_{\nu} - \partial_{\nu}\,B_{\mu}
\end{equation}

In this section instead of giving the complete formulation of the
problem we give a special solution of collision problem. We consider
the same spacetime structure as considered in the previous section
with the line element (\ref{me1}).   In the general case none of the waves
superpose due to the
nonlinearities in the field equations. On the other hand existence of two
different abelian gauge fields allow one to consider the following
type of collision
problem (such a solution does not exist with one abelian gauge field).
Consider one of the gauge fields is zero in one of the incoming
regions and the other gauge field is zero in the other incoming region.
More specifically
one of the $U(1)$ potential ($A_{\mu}$) vanishes in one of the incoming
regions
and the other $U(1)$ potential ($B_{\mu}$) vanishes in the other
region. In the interaction region we have both fields.
This implies a superposition in the gauge fields.
Such an  assumption simplifies the field equations considerably \cite{GUR}.

The reduced field equations are as follows

\begin{eqnarray}
U_{uv}-U_{u}\,U_{v}=0    \label{s30}\\
2V_{uv}-U_{u}V_{v}-U_{v}V_{u}=0
\end{eqnarray}

\begin{equation}
-2M_{u}U_{u}-2\,U_{uu}+U_{u}^{2}+
(1+{8 \over a^{2}})\,V_{u}^{2}+4\,\kappa^{2}\,B_{u}^{2}\,e^{U}=0 \label{an1}
\end{equation}

\begin{equation}
-2M_{v}U_{v}-2U_{vv}+U_{v}^{2}+
(1+{8 \over a^{2}})\,V_{v}^{2}+4\,\kappa^{2}\,A_{v}^{2}\,e^{U}=0
\label{an2}
\end{equation}

\noindent
Depending upon the choices of the $U(1)$ potentials we find the dilaton
field $\psi$ accordingly. We have two distinct cases

\noindent
{\bf Case 1: b=a}\, We have two subcases (in each cases we assume that $a$
is different then zero).\\
{\bf (1.a)}: $\psi={1 \over a}\,V$, $A_{\mu}=(0,0,0,A(v))$ ,
$B_{\mu}=(0,0,0,B(u))$.
The field equations are given above (\ref{s30}-\ref{an2})

\noindent
{\bf (1.b)}: $\psi={1 \over a}\,V $, $A_{\mu}=(0,0,A(u),0)$ ,
$B_{\mu}=(0,0,B(v),0)$.
The field equations are exactly the same as in case (1.a)
if $A$ and $B$ are interchanged in the equations (\ref{an1}) and (\ref{an2}).

\noindent
{\bf Case 2: b=-a} We have again two subcases\\
{\bf (2.a)}: $\psi={1 \over a}\,V $, $A_{\mu}=(0,0,0,A(v))$ ,
$B_{\mu}=(0,0,B(u),0)$. The field equations are exactly the
same as in case (1.a).

\noindent
{\bf (2.b)}: $\psi=-{1 \over a}\,V $, $A_{\mu}=(0,0,A(u),0)$ ,
$B_{\mu}=(0,0,0,B(v))$. The field equations are exactly the
same as in case (1.b).

\noindent
The solutions of the equations (\ref{s30} - \ref{an2}) are given
as follows \cite{sz2}

\begin{equation}
e^{-U}=f(u)+g(v) \label{s40}
\end{equation}

\begin{equation}
V={1 \over \sqrt{f+g}}(R+S) \label{s41}
\end{equation}

\begin{eqnarray}
R=\int^{1 \over 2}_{f}\,P_{-{1 \over 2}}\,\left(1+{2(\xi-f)({1 \over 2}-g)
\over (\xi+{1 \over 2})(f+g)} \right)\,
{d \over d\xi}\left[\sqrt{{1 \over 2}+\xi}\, V_{2}(\xi)\right]\,d\xi \\
S= \int^{1 \over 2}_{g}\,P_{-{1 \over 2}}\,\left(1+{2(\eta-g)({1 \over 2}-f)
\over (\eta+{1 \over 2})(f+g)} \right)\,
{d \over d\eta} \left[\sqrt{{1 \over 2}+\eta}\, V_{3}(\eta)\, \right]\,d\eta \,
\end{eqnarray}

\noindent
where $f$ and $g$ are functions of $u$ and $v$ respectively ,
$P_{-{1 \over 2}}$
is the Legendre function of order $-{1 \over 2}$. These
functions are determined from the data . In the incoming regions
we have $f={1 \over 2}$ ($u \le 0$) and $g={1 \over 2}$  ($v \le 0$)
where

\begin{equation}
f= e^{U_{2}}- {1 \over 2} ~ , ~ g= e^{U_{3}}- {1 \over 2}
\end{equation}

\noindent
The functions $V_{2}(u)$ and $V_{3}(v)$ are the data for the function
$V(u,v)$. The solutions may be summarized as follows. Here we are giving
the case (1.a) explicitly. The other cases can be given easily
by correct identifications.

\noindent
{\bf Second region}: $v \le 0 , u > 0$ or $g={1 \over 2}$

\noindent
The dilaton field $\psi_{2}={1 \over a}\,V_{2}$ , the gauge potentials are
given as
$A_{\mu}=0$ and $B_{\mu}=(0,0,0,B(u))$. The only field equation is given
by

\begin{equation}
-2M_{2,u}U_{2,u}-2\,U_{2,uu}+U_{2,u}^{2}+
(1+{8 \over a^{2}})\,V_{2,u}^{2}+4\,\kappa^{2}\,B_{u}^{2}\,e^{U_{2}}=0
\label{s32}
\end{equation}

\noindent
{\bf Third region}: $u \le 0 , v > 0$ or $f={1 \over 2}$

\noindent
The dilaton field $\psi_{3}={1 \over a}\,V_{3}$ , the gauge potentials are
given as
$A_{\mu}=(0,0,0,A(v))$ and $B_{\mu}=0$. The only field equation is given
by

\begin{equation}
-2M_{3,v}U_{3,v}-2U_{3,vv}+U_{3,v}^{2}+
(1+{8 \over a^{2}})\,V_{3,v}^{2}+4\,\kappa^{2}\,A_{v}^{2}\,e^{U_{3}}=0
\label{s34}
\end{equation}

\noindent
{\bf Fourth region}: $u > 0 , v > 0$

\noindent
where the exact solutions of $U(u,v)$ and $V(u,v)$ are given in (\ref{s40})
and (\ref{s41}).
The dilaton field $\psi(u,v)={1 \over a}\,V(u,v)$ , the gauge potentials are
given as
$A_{\mu}=(0,0,0,A(v))$ and $B_{\mu}=(0,0,0,B(u))$. The field equations
to be solved are (\ref{an1}) and (\ref{an2}). Given the data ($V_{2}(u) ,
V_{3}(v)$) one finds the function $V(u,v)$ from the integral formula
(\ref{s41}).
Given the data ($V_{2}(u) ,V_{3}(v)$) and ($A(v),B(u)$) one integrates
the function $M(u,v)$ from (\ref{an1}) and (\ref{an2}).

\noindent
A simple exact solution to the above problem is given as follows

\begin{equation}
V=m_{1}\, \tanh^{-1}\,\left({{1 \over 2}-f \over {1 \over 2}+g} \right)^{1
\over 2}
  +m_{2}\, \tanh^{-1}\,\left({{1 \over 2}-g \over {1 \over 2}+f} \right)^{1
\over 2}
\end{equation}

\noindent
with

\begin{eqnarray}
V_{2}=m_{1}\, \tanh^{-1}\,({1 \over 2}-f)^{1 \over 2} \label{ss0}\\
V_{3}=m_{2}\, \tanh^{-1}\,({1 \over 2}-g)^{1 \over 2} \label{ss1}
\end{eqnarray}

\noindent
where $m_{1}$ and $m_{2}$ are arbitrary constants. In the general case
the initial data is loaded on the functions $f$ and $g$. The determination
of these functions is important in the integration of the function $M$.
We find this function
by following two different approaches. This means that we have two different
solutions for two different data.

\vspace{0.5cm}

\noindent
{\bf First Solution :} \,\,The functions $f$ and $g$ are given by

$$f(u)= {1 \over 2}-s_{1}\, u^{n_{1}}\, \theta(u)~~ ,~~
g(v)= {1 \over 2}-s_{2}\, v^{n_{1}}\, \theta(v)$$

\noindent
where $n_{1}$ and
$n_{2}$ are positive integers ($\ge 2$). This is not the complete data but the
function $M(u,v)$ can be found as

\begin{eqnarray}
2\,M=(1-{b \over 4}\,(m_{1}+m_{2})^{2})\, \ln (f+g)+{b \over 4}\, \left[
m_{1}^{2}\, \ln ({1 \over 2}+g)+m_{2}^{2}\, \ln ({1 \over 2}+f) \right]+
\nonumber \\
{b \over 2}\, m_{1}\, m_{2}\, \ln({1 \over 2}+2\,f\,g+
{1 \over 2}\,\sqrt{(1-4\,f^{2})(1-4\,g^{2})})
-4\, \kappa^{2}\,\left( \int_{1 \over 2}^{f}\,B_{\xi}^{2}\,d\xi+
\, \int_{1 \over 2}^{g}\,A_{\eta}^{2}\,d\eta \right) \nonumber
\end{eqnarray}

\noindent
In the incoming regions we have

\begin{eqnarray}
2\,M_{2}=(1-{b \over 4}\,m_{1}^{2})\, \ln ({1 \over 2}+f)
-4\, \kappa^{2}\, \int_{1 \over 2}^{f}\,B_{\xi}^{2}\,d\xi\\
2\,M_{3}=(1-{b \over 4}\,m_{2}^{2})\, \ln ({1 \over 2}+g)
-4\, \kappa^{2}\, \int_{1 \over 2}^{g}\,A_{\eta}^{2}\,d\eta
\end{eqnarray}
\noindent
where the last two integrals in above expression are due to the
initial values of the gauge fields on the null hyperplanes which
are left arbitrary and

$$b=1+{8 \over a^{2}} ~~ , ~~ b\,m_{i}^{2}=8\,(1-{1 \over n_{i}})$$

\noindent
with $i=1,2$. As far as the singularity structure is considered
our solution given above looks like the vacuum Einstein solutions
given by Szekeres \cite{sz2}. They all suffer from a future closing
spacetime singularity at $f+g=0$.

\vspace{0.5cm}

\noindent
{\bf Second Solution:}\,\, The functions $f$ and $g$ are determined
by the following equations

\begin{equation}
{2\,f_{uu} \over f_{u}^{2}}
={1 \over {1 \over 2}+f}-b\,({1 \over 2}+f)\,\left({dV_{2} \over
df}\right)^{2}-
4\, \kappa^{2}\, \left({dB \over df}\right)^{2} \label{ff1}
\end{equation}

\begin{equation}
{2\,g_{vv} \over g_{v}^{2}}
={1 \over {1 \over 2}+g}-b\,({1 \over 2}+g)\,\left({dV_{3} \over
dg}\right)^{2}-
4\, \kappa^{2}\, \left({dA \over dg}\right)^{2} \label{ff2}
\end{equation}

\noindent
where $V_{2}$ and $V_{3}$ are given in (\ref{ss0})and (\ref{ss1}).
Then the function $M(u,v)$ is found as

\begin{eqnarray}
2\,M&=&\left[1-b\,{(m_{1}+m_{2})^{2} \over 4} \right]\, \ln(f+g)
+\left[-1+b\,{m_{1}^{2}+m_{2}^{2} \over 4} \right]\, \ln \left(({1 \over 2}+f)
({1 \over 2}+g) \right) \nonumber \\
 & &+b\,{m_{1}\,m_{2} \over 2}\, \ln \left({1 \over 2}+2fg+ {1 \over 2}\,
\sqrt{(1-4f^{2})(1-4g^{2}} \right)
\end{eqnarray}

\noindent
The function $M$ in the incoming regions vanish ($M_{2}=M_{3}=0$).
Hence given the functions $A(g)$ and $B(f)$, we determine
the functions $f$ and $g$ through (\ref{ff1}) and (\ref{ff2})
in terms of $u$ and $v$. This completes the determination
of the metric in the fourth region. For different set of
functions ($A(g),B(f)$) we have different solutions.

\noindent
When the gauge potentials $A$ and $B$ go to zero and the dilaton
couping constant becomes larger then both of the above solutions
approach to the Szekeres solutions (\cite{sz2}).
For all of these solutions the surface $f+g=0$ is singular.

\section{Conclusion}

We have given exact solutions of the colliding plane waves in the
Einstein Maxwell Dilaton Gravity theories. Although
the exact solutions we obtained in this work differ from the solutions
of the vacuum Einstein and Einstein Maxwell theories,
the singularity structures of the solutions of these different theories
look the same. In this work we have studied the collision of plane
waves in four dimensions. Higher dimensional plane waves when
dimensionally reduced (with some duality transformations) lead to
the extreme black hole solutions in four dimensions. In this respect
it is perhaps more interesting to investigate the colliding gravitational
plane waves in higher dimensions. This will be the subject of forthcoming
communication.

\section{Appendix}

In Maxwell theory , due to the linearity the solution in the
interaction region is just the superposition of the plane wave
solutions in the secod and third regions. In Einstein theory
such a superposition is not allowed and hence to find exact solutions
(solution of the charcteristic initial value problem) is not possible yet.
In this appendix we consider the collision of the
Maxwell-Dilaton Plane Waves which shares the similar difficulties
of the Einstein theory. The lagrangian of the coresponding theory
is

\begin{equation}
L=\left[{2\over \kappa^{2}}(\nabla \psi)^{2}+
{1\over 4}\ e^{-a \psi}F^{2}\right].
\end{equation}

\noindent
where $a$ is the dilaton coupling constant and the spacetime
metric is flat in all regions.
Here we kept the
the constant $\kappa$ which may be set equal to unity.
The field equations are

\begin{equation}
\nabla_{\mu} (e^{-a \psi } F^{\mu \nu}) = 0,  \label{ap1}
\end{equation}

\begin{equation}
\partial_{\mu} ({\sqrt -g}\, g^{\mu \nu}\, \partial_{\nu} \psi)+
{\kappa^{2}\, a\, {\sqrt -g} \over 16} e^{-a \psi} F^2 = 0,  \label{ap2}
\end{equation}

\noindent
With the choice $A_{\mu}=(0,0,A,0)$, where $A=A(u,v)$  and
dilaton field $\psi=\psi(u,v)$, the field equations turn out to be

\begin{equation}
a\psi_{u} A_{,v}+a \psi_{v} A_{,u}-2 A_{uv}=0 \label{ap3}
\end{equation}

\noindent
and

\begin{equation}
\psi_{uv} +{a \kappa^{2}\over 8} e^{-a \psi} A_{,u} A_{,v}=0. \label{ap4}
\end{equation}

\noindent
These equations are the real and imaginary parts of the following
Ernst equation

\begin{equation}
Re(\epsilon)\, \nabla^{2}\, \epsilon =
\nabla \epsilon\, \nabla \epsilon  \label{dok}
\end{equation}

\noindent
where differential operators in (\ref{dok}) are defined with
respect to the metric given by $ds^{2}=2\,du\,dv$ and

\begin{equation}
\epsilon=e^{{1 \over 2}a \psi}+i{a \kappa \over 4}\,A
\end{equation}

\noindent
This can be rewritten as

\begin{equation}
\nabla (g^{-1}\, \nabla g)=0  \label{ap6}
\end{equation}

\noindent
where

\begin{equation}
g={2\over \epsilon +\overline{\epsilon}}\left[
\begin{array}{cc}
1&{i\over 2} (\epsilon -\overline{\epsilon}) \\
{i\over 2} (\epsilon -\overline{\epsilon})&\epsilon \overline{\epsilon}
\end{array}
\right]
\end{equation}

\noindent
Eq.(\ref{ap6}) is the two dimensional sigma model equation on
$SU(2)/U(1)$. Although the complete solution of (\ref{dok})
is not known yet its integrability has been shown
long time ago \cite{ZAK}. The soliton solutions and many intersting
properties are known.

\vspace{0.5cm}

\noindent
We would like to thank to TUBITAK (Scientific and Technical
Research Council of Turkey) and TUBA (Turkish Academy of Sciences)
for their partial supports to this work. MG is an associate member
of TUBA.


\end{document}